\documentclass{llncs}

\usepackage{amsmath}
\usepackage{amssymb}
\usepackage[english]{babel}
\usepackage{xcolor}
\usepackage{graphicx}
\usepackage{pgfplots}
\usepackage{textcomp}
\usepackage{listings}
\usepackage{multirow}
\usepackage{hyperref}
\usepackage{float}
\floatstyle{plain}
\newfloat{Listing}{htb}{prg}
\usepackage[labelsep=period,labelfont=bf]{caption}

\newcommand{\figref}[1]{Figure~\ref{#1}}
\newcommand{\tabref}[1]{Table~\ref{#1}}
\newcommand{\secref}[1]{Section~\ref{#1}}
\newcommand{\lstref}[1]{Listing~\ref{#1}}

\newcommand{\scoop}[1]{\lstinline[language={Eiffel},basicstyle=\normalsize]{#1}}
\newcommand{\sscoop}[1]{\lstinline[language={Eiffel},basicstyle=\small]{#1}}
\begin{document}

\title{Handling Parallelism in a Concurrency Model}
\author{Mischael Schill \and Sebastian Nanz \and Bertrand Meyer}
\institute{ETH Zurich, Switzerland \\ \email {firstname.lastname@inf.ethz.ch}}
\maketitle

\begin{abstract}
Programming models for concurrency are optimized for dealing with
nondeterminism, for example to handle asynchronously arriving
events. To shield the developer from data race errors effectively,
such models may prevent shared access to data altogether. However,
this restriction also makes them unsuitable for applications that
require data parallelism. We present a library-based approach for
permitting parallel access to arrays while preserving the
safety guarantees of the original model. When applied to SCOOP, an
object-oriented concurrency model, the approach exhibits a negligible
performance overhead compared to ordinary threaded implementations of
two parallel benchmark programs.
\end{abstract}

\section{Introduction}
\label{sec:introduction}

Writing a multithreaded program can have a variety of very different motivations~\cite{Okur:2012:DUP:2393596.2393660}. Oftentimes, multithreading is a functional requirement: it enables applications to remain responsive to input, for example when using a graphical user interface. Furthermore, it is also an effective program structuring technique that makes it possible to handle nondeterministic events in a modular way; developers take advantage of this fact when designing reactive and event-based systems. In all these cases, multithreading is said to provide \emph{concurrency}. In contrast to this, the multicore revolution has accentuated the use of multithreading for improving performance when executing programs on a multicore machine. In this case, multithreading is said to provide \emph{parallelism}. 

Programming models for multithreaded programming generally support either concurrency or parallelism. For example, the Actor model~\cite{Agha:1986:AMC:7929} or Simple Concurrent Object-Oriented Programming (SCOOP)~\cite{meyer:1997:oosc,nienaltowski:2007:SCOOP} are typical concurrency models: they are optimized for coordination and event handling, and provide safety guarantees such as absence of data races. Models supporting parallelism on the other hand, for example OpenMP~\cite{Dagum:1998:OIA:615255.615542} or Chapel~\cite{Chamberlain:2007:PPC:1286120.1286123}, put the emphasis on providing programming abstractions for efficient shared memory computations, typically without addressing safety concerns. 

While a separation of concerns such as this can be very helpful, it is evident that the two worlds of concurrency and parallelism overlap to a large degree. For example, applications designed for concurrency may have computational parts the developer would like to speed up with parallelism. On the other hand, even simple data-parallel programs may suffer from concurrency issues such as data races, atomicity violations, or deadlocks. Hence, models aimed at parallelism could benefit from inheriting some of the safety guarantees commonly ensured by concurrency models. 

This paper presents a library-based approach for parallel processing of shared-memory arrays within the framework of a concurrency model. To achieve this, the data structure is extended with features to obtain \emph{slices}, i.e.\ contiguous data sections of the original data structure. These data parts can be safely used by parallel threads, and the race-freedom guarantee for the original data structure can be preserved.

The approach is applied to SCOOP~\cite{meyer:1997:oosc,nienaltowski:2007:SCOOP}, a concurrency model implemented on top of the object-oriented language Eiffel~\cite{eiffel}. A performance evaluation using two benchmark programs (parallel Quicksort and matrix multiplication) shows that the approach is as fast as using threads, and naturally outperforms the original no-sharing approach. While SCOOP lends itself well to our approach, the basic idea can be helpful for providing similar extensions to Actor-based models.

The remainder of the paper is structured as follows. \secref{sec:problem-statement} describes the problem and the rationale of our approach. \secref{sec:solution} presents the slicing technique. \secref{sec:evaluation} provides the results of the performance evaluation. \secref{sec:related-work} describes related work and \secref{sec:conclusion} concludes with thoughts on future work.

\section{Performance issues of race-free models}
\label{sec:problem-statement}

To help conquer the complexity of nondeterministic multithreading, programming models for concurrency may provide safety guarantees that exclude common errors by construction. In Erlang~\cite{Armstrong:1996:CPE:229883} for example, a language famous for implementing the Actor model~\cite{Agha:1986:AMC:7929}, there is no shared state among actors; hence the model is free from data races. 

In a similar manner, the object-oriented concurrency model SCOOP~\cite{meyer:1997:oosc,nienaltowski:2007:SCOOP} does not allow sharing of memory between its computational entities, called \emph{processors} (an abstraction of threads, processes, physical cores etc). More specifically, every object in SCOOP belongs to exactly one processor and only this processor has access to the state of the object. A processor can however be instructed to execute a call on behalf of another processor, by adding the call to the processor's \emph{request queue}. Also this regime offers protection from data races.

Unfortunately, the strict avoidance of shared memory has severe performance disadvantages when trying to parallelize various commonplace computational problems. As an example, \lstref{lst:quicksort:scoop} shows an in-place Quicksort algorithm written in SCOOP. Every time the array is split, a new worker is created to sort its part of the array. The workers \scoop{s1} and \scoop{s2} and the array \scoop{data} are denoted as \scoop{separate}, i.e.\ they reference an object that may belong to another processor. By creating a separate object, a new processor is spawned. Calls to a separate object are only valid if the processor owning the separate object is \emph{controlled} by the current processor, which is guaranteed if the separate object appears in the argument list, hence the \scoop{separate_sort}, \scoop{get}, and \scoop{swap} features. Each processor can only be controlled by one other processor at a time, thereby ensuring freedom from data races.

\begin{Listing}[t]
\begin{lstlisting}[basicstyle=\scriptsize]
data: separate ARRAY[T]	
lower, upper: INTEGER

make (d: separate ARRAY[T]; n: INTEGER)
	do 
		if n > 0 then
			lower := d.lower
			upper := d.lower + n - 1
		else
			upper := d.upper
			lower := d.upper + n + 1
		end
		data := d
	end
	
sort
	local i, j: INTEGER; s1, s2: separate SORTER[T]
	do
		if upper > lower then
			pivot := get (data, upper)
			from i := lower; j := lower	until i = upper loop
				if get(data, i) < pivot then
					swap(data, i, j)
					j := j + 1
				end
				i := i + 1
			end
			swap (data, upper, j)
			create s1.make (data, j - lower)
			create s2.make (data, j - upper)
			separate_sort(s1, s2)
		end
	end
	
get (d: separate ARRAY[T]; index: INTEGER): T
	do Result := data[index] end
			
swap (data: separate ARRAY[T]; i, j: INTEGER)
	local tmp: T do tmp := d[i]; d[i] := d[j]; d[j] := tmp end
				
separate_sort (s1, s2: separate SORTER[T])
	do s1.sort; s2.sort end
\end{lstlisting}
\caption{SORTER: In-place Quicksort in SCOOP}
\label{lst:quicksort:scoop}
\end{Listing}


The execution of this example exhibits parallel slowdown: a sequential version outperforms the algorithms for most workloads. This has two main reasons:
\begin{enumerate}
\item Every call to the data array involves adding the call to the request queue, removing the call from the request queue, and sending back the result; this creates a large communication overhead.
\item Only one of the workers at a time can execute the \scoop{get} and \scoop{swap} features on the array because they require control of the processor handling the array; this serialization prohibits the algorithm from scaling up.
\end{enumerate}

The same issues occur in a broad range of data-parallel algorithms using arrays. Efficient implementations of such algorithms are impossible in race-protective concurrency models such as SCOOP, which is unacceptable. Any viable solution to the problem has to get rid of the communication overhead and the serialization. There are two general approaches to this problem:

\begin{enumerate}
\item Weaken the concurrency model to allow shared memory without race protection, or interface with a shared memory language. The programmers are responsible to take appropriate synchronization measures themselves.
\item Enable shared memory computations, but hide it in an API that preserves the race-freedom guarantees of the concurrency model.
\end{enumerate}
The first approach shifts the burden to come up with solutions for parallelism to the programmer. Unfortunately, it also forfeits the purpose of race-protection mechanisms in the first place. Still, it is the most prominent approach taken. This paper presents a solution based on the second approach, in particular offering both race protection and shared memory performance. 

%

\section{Array slicing}
\label{sec:solution}

To allow the implementation of efficient parallel algorithms on arrays, the following two types of array manipulation have to be supported:
\begin{itemize}
\item \emph{Parallel disjoint access:} Each thread has read and write access to disjoint parts of an array.
\item \emph{Parallel read:} Multiple threads have read-only access to the same array.
\end{itemize}
%
%
The array slicing technique presented in this section enables such array manipulations by defining two data structures, \emph{slices} and \emph{views}, representing parts of an array that can be accessed in parallel while maintaining race-freedom guarantees. 
\begin{description}
\item[Slice] Part of an array that supports read and write access of single threads.
\item[View] Proxy that provides read-only access to a slice while preventing modifications to it.
\end{description}

In the following we give a specification of the operations on slices and views.

\subsection{Slices}
\label{sec:slices}

\begin{table}[t]
\centering
\begin{tabular}{lp{5.6cm}}
\multicolumn{2}{l}{\textit{Creation procedures (constructors)}} \\
\hspace{1em}\sscoop{make(n: INTEGER)}                      & Create a new slice with a capacity of \sscoop{n} \\
\hspace{1em}\sscoop{slice_head(slice: SLICE; n: INTEGER)} & Slice off the first \sscoop{n} entries of \sscoop{slice} \\
\hspace{1em}\sscoop{slice_tail(slice: SLICE; n: INTEGER)} & Slice off the last \sscoop{n} entries of \sscoop{slice} \\
\hspace{1em}\sscoop{merge(a, b: SLICE)} & Create a new slice by merging \sscoop{a} and \sscoop{b} \\

\multicolumn{2}{l}{\textit{Queries}} \\
\hspace{1em}\sscoop{item(index: INTEGER): T}  & Retrieve the item at \sscoop{index} \\
\hspace{1em}\sscoop{indexes: SET[INTEGER]}    & Indexes of this slice \\
\hspace{1em}\sscoop{lower: INTEGER}           & Lowest index of the index set \\
\hspace{1em}\sscoop{upper: INTEGER}           & Highest index of the index set \\
\hspace{1em}\sscoop{count: INTEGER}           & Number of indexes: \sscoop{upper - lower + 1} \\
\hspace{1em}\sscoop{is_modifiable: BOOLEAN}   & Whether the array is currently modifiable, i.e.\ \sscoop{readers = 0} \\
\hspace{1em}\sscoop{readers: INTEGER}         & The number of views on the slice \\

\multicolumn{2}{l}{\textit{Commands}} \\
\hspace{1em}\sscoop{put(value: T; index: INTEGER): T} & Store \sscoop{value} at \sscoop{index} \\
\multicolumn{2}{l}{\textit{Commands only accessible to slice views}} \\
\hspace{1em}\sscoop{freeze}                           & Notifies the slice that a view on it is created by incrementing \sscoop{readers}\\
\hspace{1em}\sscoop{melt}                             & Notifies the slice that a view on it is released by decrementing \sscoop{readers}\\

\multicolumn{2}{l}{\textit{Internal state}} \\
\hspace{1em}\sscoop{area: POINTER}            & Direct unprotected memory access \\
\hspace{1em}\sscoop{base: INTEGER}            & The offset into memory \\
\end{tabular}
\vspace{1ex}
\caption{API for slices}
\label{tab:slices}
\end{table}


Slices enable parallel usage patterns of arrays is where each thread works on a disjoint part of the array. The main operations on slices are defined as follows: 
\begin{description}
\item[Slicing] Creating a slice of an array transfers some of the data of the array into the slice. If shared memory is used, the transfer can be done efficiently using aliasing of the memory and adjusting the bounds of the original array.
\item[Merging] The reverse operation of slicing. Merging two slices requires them to be adjacent to form an undivided range of indexes. The content of the two adjacent slices is transferred to the new slice, using aliasing if the two are also adjacent in shared memory.
\end{description}
Based on this central idea, an API for slices can be defined as in \tabref{tab:slices}. Note that we use the letter \scoop{T} to refer to the type of the array elements. 
After creating a new slice using \scoop{make}, the slice can be used like a regular array using \scoop{item} and \scoop{put} with the \scoop{indexes} ranging from \scoop{lower} to \scoop{upper}, although modifying it is only allowed if \scoop{is_modifiable} is true, which is exactly if \scoop{readers} is zero. 
Internally, the attribute \scoop{area} is a direct pointer into memory which can be accessed like a 0-based array. The \scoop{base} represents the base of the slice, which is usually 1 for Eiffel programs, but may differ when a merge results in a copy. The operations \scoop{freeze} and \scoop{melt} increment and decrement the \scoop{readers} attribute which influences \scoop{is_modifiable} and are used by views (see section \ref{sec:views}). 

\paragraph{Slicing.} Like any other object, a reference to the slice can be passed to other processors. A processor having a reference to a slice can decide to create a new slice by slicing from the lower end (\scoop{slice_head}) or upper end (\scoop{slice_tail}). By doing this, the original slice transfers data to the new slice by altering the bounds and referencing the same memory if possible. Freedom of race conditions is ensured through the exclusive access to the disjoint parts.

\lstref{lst:slicing} shows an implementation of the \scoop{slice_head} creation procedure, taking advantage of shared memory. It copies the \scoop{lower} bound, the \scoop{base} and the memory reference of the slice \scoop{a_original}. It also sets the upper bound according to the size \scoop{n} of the new slice and increases the lower bound of the original by \scoop{n}.

We use Eiffel for our implementation. Eiffel provides preconditions and postconditions, which we use to make sure only modifiable arrays are altered.

\begin{Listing}[t]
\begin{lstlisting}[basicstyle=\scriptsize]
slice_head (n: INTEGER; a_original: separate SLICE[T])
	require -- Precondition
		within_bounds: n > 0 and n <= a_original.count
		a_original.is_modifiable
	do
		lower := a_original.lower
		upper := a_original.lower + n - 1
		base := a_original.base
		area := a_original.area
		a_original.lower := a_original.lower + n
	ensure -- Postcondition
		a_original.count = old a_original.count - n
		a_original.lower = old a_original.lower + n
		a_original.upper = old a_original.upper
		lower = old a_original.lower
		upper = a_original.lower - 1
		count = n
		-- "forall i in indexes : item(i) = old a_original.item(i)"
	end
\end{lstlisting}
\caption{Slicing}
\label{lst:slicing}
\end{Listing}

\paragraph{Merging.} If a processor has two \emph{adjacent} slices (the lower index of the one equals the upper index of the other plus one), calling \scoop{merge} creates a new combined slice. This transfers all the data from the old slices to the new one, making the old ones empty. If the two slices are located next to each other in memory, the transfer simply adjusts the bounds; otherwise, it copies the data into a new slice. 

The implementation of merging (see \lstref{lst:merging}) sets the bounds according to the arguments. It then checks whether the two parts are actually next to each other in memory by checking whether the \scoop{area} and the \scoop{base} are the same. In this case, it copies the base and the memory reference. Otherwise, it allocates new memory and copies all the data of the two arguments. In the end, it empties the two arguments, setting their \scoop{count} to 0 by making \scoop{lower} = 1 and \mbox{\scoop{upper} = 0}.

\begin{Listing}
\begin{lstlisting}[basicstyle=\scriptsize]
merge (a_one, a_another: separate SLICE[T])
	require
		a_one.is_modifiable
		a_another.is_modifiable
		one.is_adjacent (a_another)
	do
		lower := a_one.lower.min(a_another.lower)
		upper := a_another.upper.max(a_one.upper)
		if  a_one.area = a_another.area and a_one.base = a_another.base then
			area := a_one.area
			base := a_one.base
		else
			base := lower
			-- "Copy data from the a_one and a_another to area"
		end
		a_another.empty; a_one.empty
	ensure
		lower = old a_one.lower.min(a_another.lower)
		upper = old a_one.upper.max(a_another.upper)
		a_one.count = 0 and	a_another.count = 0
		-- "forall i in old a_one.indexes : item(i) = old a_one.item(i)"
		-- "forall i in old a_another.indexes : item(i) = old a_another.item(i)"
	end
\end{lstlisting}
\caption{Merging}
\label{lst:merging}
\end{Listing}

\paragraph{Strategies for slicing.} The most common choice for disjoint index subsets are sets with contiguous indexes. Those subsets can be identified by their lower and upper index and resemble a normal array. A rarer case is to create the disjoint subsets according to another principle. This warrants a different implementation, which is possible by using inheritance. However, current cache architectures limit the usefulness of slices with a size smaller than a cache line. 


\subsection{Views}
\label{sec:views}

Views enable read-only access on arrays. The main operations on views are defined as follows: 

\begin{description}
\item[Viewing] Creating a view from a slice copies the bounds and the memory reference into the view. The original slice is no longer modifiable.
\item[Releasing] The reverse operation of viewing. If no other views on the same slice exist, it is modifiable again. Also, the view is no longer usable.
\end{description}

\begin{table}[t]
\centering
\begin{tabular}{l@{\hspace{2cm}}p{5cm}}
\multicolumn{2}{l}{\textit{Creation procedures (constructors)}} \\
\hspace{1em}\sscoop{make(slice: SLICE)}      & Create a new view on \sscoop{slice} \\

\multicolumn{2}{l}{\textit{Queries}} \\
\hspace{1em}\sscoop{original: SLICE[T]}      & Slice this view references \\
\hspace{1em}\sscoop{indexes: SET[INTEGER]}   & Indexes of this view \\
\hspace{1em}\sscoop{item(index: INTEGER): T} & Retrieve the item at index \\
\hspace{1em}\sscoop{lower: INTEGER}          & Lowest index of the index set \\
\hspace{1em}\sscoop{upper: INTEGER}          & Highest index of the index set \\

\multicolumn{2}{l}{\textit{Commands}} \\
\hspace{1em}\sscoop{free}                    & Disconnects the view from the slice \\

\multicolumn{2}{l}{\textit{Internal state}} \\
\hspace{1em}\sscoop{area: POINTER}            & Direct unprotected memory access \\
\hspace{1em}\sscoop{base: INTEGER}            & Offset into memory \\
\end{tabular}
\vspace{1ex}
\caption{API for slice views}
\label{tab:sliceviews}
\end{table}

The API for views is shown in \tabref{tab:sliceviews}. A processor is able to read the slice in parallel by creating a view using the \scoop{make} creation procedure. The original slice is then available as the \scoop{original} query. This also prevents all further modifications of the array unless the view is released with the \scoop{free} procedure. All the other features of views behave exactly like their counterparts in the slices.

\paragraph{Viewing.} Creating a view basically copies the bounds and the memory reference into the view. By increasing the view counter (\scoop{readers}) using the \scoop{freeze} operation of the slice \scoop{a_original}, the original slice is no longer modifiable (see \lstref{lst:viewing}). By calling \scoop{free} on a view, the view loses its reference to the memory of the slice and the original slice is notified through \scoop{melt} that there is one less \scoop{reader}.

\begin{Listing}[t]
\begin{lstlisting}[basicstyle=\scriptsize]
make (a_original: separate SLICE[T])
	do
		a_original.freeze
		original := a_original
		lower := a_original.lower
		upper := a_original.upper
		base := a_original.base
		area := a_original.area
	ensure
		lower = a_original.lower
		upper = a_original.upper
		not a_original.is_modifiable
		-- "forall i in indexes : item(i) = a_original.item(i)"
	end
\end{lstlisting}
\caption{Viewing}
\label{lst:viewing}
\end{Listing}

\paragraph{Releasing.} The \scoop{free} procedure redirects the \scoop{area} to 0 and sets \scoop{lower} to 1 and \scoop{upper} to 0. Therefore no access is possible at any index. In addition, the number of readers of the original decremented by a call to \scoop{melt}. This causes the original to be modifiable again if the number of readers falls to zero. Because of its simplicity, the code is omitted. 

%
%

\section{Performance evaluation}
\label{sec:evaluation}

To assess the performance of our approach, we apply it to two benchmark problems: to determine how well our approach works in a divide-and-conquer scenario, we choose a parallel in-place Quicksort algorithm; to determine the raw performance, we use parallel matrix multiplication. In both cases, the extension of SCOOP with the slicing technique is compared with implementations in Eiffel using only threads and without synchronization except a join at the end.

For the performance tests we use a server with four 8-core Intel Xeon E7-4830 processors and 256 GB of RAM. We ran every program 20 times and report the mean value of the running times 
in \tabref{tab:measurements}. 
The source code of the benchmarks is available online\footnote{\url{https://bitbucket.org/mischaelschill/array-benchmarks}} as well as the support for slicing in SCOOP\footnote{\url{https://bitbucket.org/mischaelschill/scoop-library}}.
In the following we discuss both benchmarks and their results in detail.

\begin{table}[t]
  \centering
  \def\arraystretch{1.2}
  \begin{tabular}{l@{\ }l@{\ }|@{\ }rrrrrr}
  & & \multicolumn{6}{c}{Number of cores} \\
  & & 1 & 2 & 4 & 8 & 16 & 32 \\ 
  \hline 

  \multirow{2}{*}{Quicksort} 
    & Slicing & 157.4 & 147.1 & 81.9 & 66.4 & 59.9 & 59.2 \\ 
    & Threads & 158.6 & 145.1 & 82.8 & 68.0 & 61.5 & 59.8 \\ 
  \hline 
  \multirow{2}{*}{Matrix multiplication}
    & Slicing & 184.8 & 95.0 & 51.2 & 24.0 & 14.1 & 7.3 \\ 
    & Threads & 178.0 & 91.7 & 46.6 & 23.6 & 12.6 & 7.3 \\
  \end{tabular} 
  \vspace{2ex}
  \caption{Mean running times (in seconds)}
  \label{tab:measurements}
\end{table}

\subsection{Quicksort}
\label{sec:quicksort}

\lstref{lst:quicksort:split} shows the constructor of the Quicksort example implemented using slices instead of regular arrays (compare \secref{sec:problem-statement}). The main difference is the usage of \scoop{slice_head} and \scoop{slice_tail} instead of storing the bounds in variables. The implementation of \scoop{sort} can stay the same, although there is no need for storing the bounds and extra features for swapping and retrieving since data is no longer separate. 

\begin{Listing}
\begin{lstlisting}[basicstyle=\scriptsize]
data: SLICE[T]

make (a_data: SLICE[T]; n: INTEGER)
	do
		if n > 0 then
			create data.slice_head (a_data, n)
		else
			create data.slice_tail (a_data, -n)
		end
	end	 
\end{lstlisting}
\caption{Quicksort algorithm using slices}
\label{lst:quicksort:split}
\end{Listing}

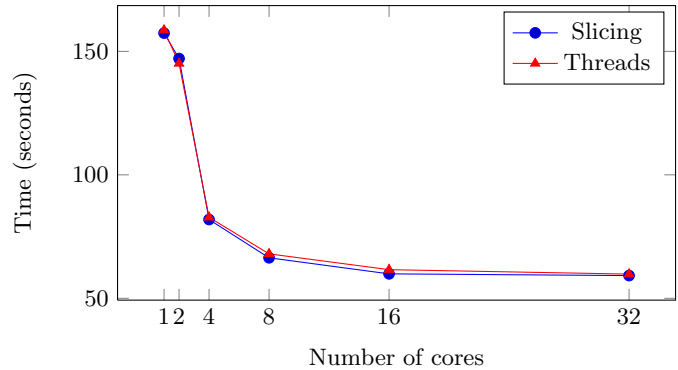
\begin{figure}[t]
  \centering

\begin{tikzpicture}
\begin{axis}[scaled ticks=false,xlabel=Number of cores,ylabel=Time (seconds),xtick={1,2,4,8,16,32},height=5.5cm,width=9cm]
\addplot
coordinates {
	(1,157.4) 
	(2,147.1)
	(4,81.9) 
	(8,66.4) 
	(16,59.9)
	(32,59.2)
};
\addlegendentry{Slicing}
\addplot[mark=triangle*,color=red]
coordinates {
	(1,158.6) 
	(2,145.1)
	(4,82.8) 
	(8,68.0) 
	(16,61.6)
	(32,59.8)
};
\addlegendentry{Threads}
\end{axis}
\end{tikzpicture}
  \vspace{-2ex}
  \caption{Quicksort: scalability}
  \label{fig:scalability-quicksort}
\end{figure}

For the performance measurement, the Quicksort benchmark sorts an array of size $10^8$, which is first filled using a random number generator with a fixed seed. The benchmarked code is similar to listing \ref{lst:quicksort:split}, but also adds a limit on the number of processors used. As evident from \figref{fig:scalability-quicksort}, the performance characteristics of the slicing technique and threading is almost identical. 

\subsection{Matrix multiplication}
\label{sec:matrix-multiplication}

\lstref{lst:matmul:split} shows a class facilitating parallel multiplication of matrices, using a two dimensional version of slices and views (\scoop{SLICE2} and \scoop{SLICE_VIEW2}, implemented in a very similar fashion to the one-dimensional version discussed in \secref{sec:slices}). The worker is created using \scoop{make} which slices off the first \scoop{n} rows into \scoop{product}. The \scoop{multiply} command actually fills the slice with the result of the multiplication of the left and right matrices. Afterwards, the views are decoupled using \scoop{free}. Dividing the work between multiple workers and merging the result is left to the client of the worker.

\begin{Listing}
\vspace{2ex}
\begin{lstlisting}[basicstyle=\scriptsize]
left, right: SLICE_VIEW2[T]
product: SLICE2[T]

make (l, r, p: separate SLICE2[T]; n: INTEGER)
	do 
		create left.make(l); create right.make(r)
		create product.slice_top (n, p) 
	end

multiply
	local k, i, j: INTEGER
	do 
		from i := product.first_row until i > product.last_row loop
			from j := product.first_column until j > product.last_column loop
				from k := left.first_column until k > left.last_column loop
					product[i, j] := product[i, j] + left[i, k] * right [k, j]
					k := k + 1
				end
				j := j + 1
			end
			i := i + 1
		end
		left.free; right.free
	end
\end{lstlisting}
\caption{Matrix multiplication worker using slices and views}
\label{lst:matmul:split}
\end{Listing}

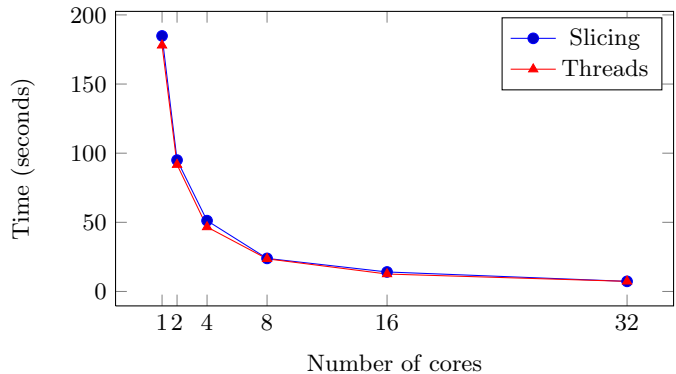
\begin{figure}[t]
  \centering

\begin{tikzpicture}
\begin{axis}[scaled ticks=false,xlabel=Number of cores,ylabel=Time (seconds),xtick={1,2,4,8,16,32},height=5.5cm,width=9cm]
\addplot
coordinates {
	(1,184.8) 
	(2,95.0)
	(4,51.2) 
	(8,23.9) 
	(16,14.1)
	(32,7.3)
};
\addlegendentry{Slicing}
\addplot[mark=triangle*,color=red]
coordinates {
	(1,178.0) 
	(2,91.7)
	(4,46.6) 
	(8,23.6) 
	(16,12.6)
	(32,7.3)
};
\addlegendentry{Threads}
\end{axis}
\end{tikzpicture}
  \vspace{-2ex}
  \caption{Matrix multiplication: scalability}
  \label{fig:scalability-matrix-multiplication}
\end{figure}

For the performance measurement, the matrix multiplication test multiplies a 2000 to 800 matrix with an 800 to 2000 matrix. \figref{fig:scalability-matrix-multiplication} shows again similar performance characteristics between slicing and threads.

%

\section{Related work}
\label{sec:related-work}

We are not aware of any programming model supporting slicing while avoiding race conditions. However, similar means to create an alias to a subset of an array's content are common in most programming languages or their standard library. For example, the standard library of Eiffel as provided by Eiffel Software~\cite{eiffel} can create subarrays. Perl~\cite{perl} has language integrated support for slicing arrays. Slices and slicing are a central feature of the Go programming language~\cite{golang}. However, these slicing solutions were not created with the intention of guaranteeing safe access: the portion of memory aliased by the new array/slice remains accessible through the original array, which can lead to race conditions if two threads access them at the same time.

Enabling many processors to access different parts of a single array is a cornerstone of data parallel programming models. OpenMP~\cite{Dagum:1998:OIA:615255.615542} is the de-facto standard for shared-memory multiprocessing. Its API offers various data parallel directives for handling the access to arrays, e.g.\ in conjunction with parallel-for loops. Threading Building Blocks~\cite{DBLP:books/daglib/0018624} is a C++ library which offers a wide variety of algorithmic skeletons for parallel programming patterns with array manipulations. Chapel~\cite{Chamberlain:2007:PPC:1286120.1286123} is a parallel programming language for high-performance computation offering concise abstractions for parallel programming patterns. Data Parallel Haskell~\cite{Chakravarty:2007:DPH:1248648.1248652} implements the model of nested data parallelism (inspired by NESL~\cite{Blelloch:1993:NND:865114}), extending parallel access also to user-defined data structures. In difference to our work, these approaches focus on efficient computation but not on safety guarantees for concurrent access, which is our starting point. 

The concept of views is an application of readers-writers locks first introduced by Courtois, Heymans and Parnas~\cite{Courtois:1971:CCL:362759.362813}, tailored to the concept of slices.

\section{Conclusion}
\label{sec:conclusion}

While programming models for concurrency and parallelism have different goals, they can benefit from each other: concurrency models provide safety mechanisms that can be advantageous for parallelism as well; parallelism models provide performance optimizations that can also be profitable in concurrent programming. In this paper, we have taken a step in this direction by extending a concurrency model, SCOOP, with a technique for efficient parallel access of arrays, without compromising the original data-race freedom guarantees of the model. An important insight from this work is that safety and performance do not necessarily have to be trade-offs: results on two typical benchmark problems show that our approach has the same performance characteristics as ordinary threading.

In future work, it would be interesting to explore the relation between models for concurrency and parallelism further, with the final goal of defining a safe parallel programming approach. In particular, programming patterns such as parallel-for, parallel-reduce, or parallel-scan could be expressed in a safe manner. In order to ascertain how this API is used by programmers, empirical studies are needed.


\subsubsection*{Acknowledgments.}
The research leading to these results has received funding from the European Research Council under the European Union's Seventh Framework Programme (FP7/2007-2013) / ERC Grant agreement no. 291389, the Hasler Foundation, and ETH (ETHIIRA).


\end{document}